\documentclass[12pt]{article}
\usepackage{graphicx}
\usepackage[utf8]{inputenc}
\usepackage{amsmath}
\usepackage{amsfonts}
\usepackage{amssymb}
\usepackage{graphicx}

\begin{document}

\begin{center}
\Large {\bf  Null Cosmic Strings: Scattering by Black Holes, Optics and Spacetime Content}
\end{center}

\bigskip
\bigskip

\begin{center}
E.A. Davydov, D.V. Fursaev, V.A. Tainov
\end{center}

\bigskip
\bigskip

\begin{center}
{\it Dubna State University \\
     Universitetskaya st. 19\\
     141 980, Dubna, Moscow Region, Russia\\

  and\\

  the Bogoliubov Laboratory of Theoretical Physics\\
  Joint Institute for Nuclear Research\\
  Dubna, Russia\\}
 \medskip
\end{center}

\bigskip
\bigskip

\begin{abstract}
Equations of motion of null cosmic strings near black holes, or other massive sources, are solved exactly in the weak field approximation. 
The stress-energy tensor of a null string in a curved spacetime is introduced and used to  show 
how scattering by black holes transforms linear and angular momenta of the string.
The corresponding recoil effect of a black hole and change of its angular momentum caused by a null cosmic string are calculated.
For a null string, its energy $\mu$  per unit length evolves along the null direction of the string trajectory. The evolution of $\mu$ is
connected with a string optical scalar $Z$. Optical properties of null strings are that their energy is concentrated on caustics, where $Z$ has poles.
String parameters $\mu$ and $Z$ capture important features of the spacetime where strings move.  
Explicit dependence of $\mu$ and $Z$ on the strain and Bondi news tensors of gravitational wave background, mass and angular momentum aspects are established, near the 
future null infinity, up to the 4-th order in expansion in an inverse null parameter in asymptotically flat spacetimes. 
\end{abstract}

\newpage

\section{Introduction}\label{intr}

Null strings are one-dimensional objects whose points move  along  trajectories of light rays, orthogonally to strings \cite{Schild:1976vq}. Null cosmic strings, if exist, generate a number of physical effects in the surrounding matter \cite{Fursaev:2017aap}, \cite{Fursaev:2018spa}, such as mutual transformations
of trajectories of massive bodies or light rays, when the string moves in between two trajectories, perturbations of the velocities of bodies resulting  in overdensities of matter, as well as shifts of energies of photons and additional  anisotropy of cosmic microwave background. 

We mention equivalent names of null cosmic strings which reflect their various properties: tensionless strings 
(since they can be viewed as tensionless limit of tensile cosmic strings \cite{Kibble:1976sj}, \cite{Vilenkin:2000jqa}) or massless strings (since they have zero rest mass per unit length). Note that physical effects of null strings are determined by a backreaction of spacetime geometry caused by nonzero energy of  strings. Like tensile strings 
null cosmic strings create holonomies of spacetime. The holonomies are null rotations which belong to a parabolic subgroup of the Lorentz group \cite{Fursaev:2017aap}. The group parameter of the holonomies is determined by the energy of the strings per unit 
length. We denote this energy by $\mu$.

From the point of view of observable effects, a  distinctive feature of null cosmic strings is in their optical properties.
Indeed, the null strings are analogous to null geodesic congruences which play an important role in General Relativity, see, e.g.
\cite{Adamo:2009vu}. This analogy has been  explored recently in \cite{Fursaev:2021xlm}. The 
trajectories (worldsheets) of null strings possess two unique physical parameters:  $\theta$, which measures the rate of expansion (contraction) of a small segment of the string along the null direction of the trajectory, and 
$\kappa$, which yields the rate of rotation of this segment in an orthogonal 2-plane. The complex spin coefficient $Z=\theta+i\kappa$, called the string scalar,  satisfies \cite{Fursaev:2021xlm} the following analog of Sachs' optical equations:
\begin{equation}\label{i.1}
\partial_\lambda Z+Z^2=-\Psi_0-\Phi_{00}~~,
\end{equation}
where $\lambda$ is an affine null parameter on the string trajectory, $\Psi_0$, $\Phi_{00}$ are invariants constructed from components of the Weyl and Ricci tensors, respectively.
As is shown in Sec. \ref{energy},  evolution of $\mu$  is connected with (\ref{i.1}):
\begin{equation}\label{i.2}
\partial_\lambda \mu + \theta \mu=0~~.
\end{equation}
Equation (\ref{i.2}) is another key feature of null cosmic strings. It indicates, in particular, that energy of a null string can 
be concentrated on certain domains of its world-sheet, in particular, on caustics,  light-like curves which trajectories 
of string segments are tangent to.

The aim of the present paper is to further study ``optical'' properties of null strings which may be important for their
experimental search. The first part of the paper is devoted to the energy of strings and their interaction  with black holes,
or other massive rotating sources, in the weak field
approximation. A similar analysis for tensile cosmic strings has been done in a number of publications, see, e.g. 
\cite{Lonsdale:1988x}, \cite{DeVilliers:1998xz}, \cite{Page:1998ya}.
In the second part we consider the evolution of string parameters $Z$, $\mu$ in asymptotically flat spacetimes,
along the lines of \cite{Fursaev:2021xlm}, to understand better effects caused on strings by the spacetime content (matter distribution and  gravitational wave background).

The paper is organized as follows.   
We start in Sec. \ref{stress} with a brief introduction to dynamics of null strings, then suggest a definition of the stress-energy tensor
(SET)
of a null string in an arbitrary gravitational background.  SET is a distribution with a support on the string world-sheet.
Since the world-sheet of a null string is degenerate, the action of the string cannot be defined, say, in the Nambu-Goto form
which is used for tensile strings. Therefore, SET cannot be derived from the action.
Our definition of SET in Sec. \ref{energy} satisfies the covariant conservation 
law, which implies important relation (\ref{i.2}), and reduces to known SET for a straight null string in Minkowsky spacetime
\cite{Fursaev:2017aap}.   Like light rays in optics, string trajectories may have caustics, where energy is concentrated.
The behaviour of $Z$ and $\mu$ near a caustic is discussed in Sec. \ref{caus}.
Based on SET,  Sec. \ref{char} introduces asymptotic Noether charges of a null string  
at past, $\mathcal{I}^{-}$, and future, $\mathcal{I}^{+}$, null infinities. 

Scattering of null strings on black holes is considered in Sec. \ref{scatt}. Explicit 
equations of string trajectories are obtained in the weak field approximation in Sec.
\ref{weak1}, and for rotating sources in Sec. \ref{angm}.
The scattering changes linear momenta of string segments at $\mathcal{I}^{+}$ with respect to $\mathcal{I}^{-}$.
Rotation of the source also generates variation of the angular momentum of the string. By conservation laws, these results 
imply that the black hole itself changes its velocity and direction of the spin in the gravitational field of the string.  
For a straight null string these effects are fairly universal: they depend only on the parameter $\mu G$ (where $G$ is the Newton coupling). As is shown in Sec. \ref{holon} the effects are due to the spacetime holonomy and are in precise agreement with results of
\cite{Fursaev:2017aap}.  Sec. \ref{opt} provides the relation between optical parameters 
$\theta_{\mathrm{in}},\kappa_{\mathrm{in}}$
at $\mathcal{I}^{-}$ and $\theta_{\mathrm{out}},\kappa_{\mathrm{out}}$ at $\mathcal{I}^{+}$. The string
evolution in the complex $Z$-plane is predictable during the scattering.

Our results show that $Z$ and $\mu$ are sensitive to spacetime content.
In Sec. \ref{AFST} we derive asymptotic form of $Z(\lambda,\tau)$ for null strings in asymptotically flat space-times
at large $\lambda$ (when $\mathcal{I}^{+}$ is approached). We use the Bondi-Sachs formalism and optical equation (\ref{i.1})
to find coefficients in expansion of $Z(\lambda,\tau)$ and string energy $\mu(\lambda,\tau)$ up to terms $\lambda^{-4}$. 
The mass aspect, the angular momentum aspects of the spacetime, as well as features of background gravitational 
radiation are encoded in $Z$, $\mu$ and can be recovered from physical effects produced by null strings.
These results extend the analysis of \cite{Fursaev:2021xlm}.
Short discussion of our results can be found in Sec. \ref{sum}.

\section{Stess-energy tensor of null strings}\label{stress}
\setcounter{equation}0
\subsection{Key elements of null string dynamics}\label{def}

A trajectory of a null string in a space-time $\cal M$ with coordinates $x^\mu$ is defined as 
$x^\mu=x^\mu(\lambda, \tau)$, where $\lambda$ and $\tau$ are real parameters. The trajectory is fixed by 
equations \cite{Schild:1976vq}:
\begin{equation}\label{1.1}
(l \cdot l)=0~~,~~~(\eta \cdot \eta )>0~~~,
\end{equation}  
\begin{equation}\label{1.2}
(l \cdot \eta)=0~~, 
\end{equation} 
\begin{equation}\label{1.3}
\nabla_l l=\beta l~~. 
\end{equation} 
where  $l^\mu\equiv x^\mu_{~,\lambda}$ and $\eta^\mu \equiv x^\mu_{~,\tau}$ are the tangent vectors, 
notation $(u \cdot v)$ stands for the scalar product of vectors $u$, $v$ in the tangent space of $\cal M$.  $\lambda$ is an affine parameter
if $\beta=0$. We assume that velocity of the string $l$ is future-directed.  $\eta$ is called the connecting vector.

Since any point of the string moves as a light ray, trajectories of null strings can be considered as a one-dimensional analogue of null geodesic congruences (NGC). Though one cannot create a one-dimensional string-like congruence of light rays,
the cross-section of such NGC cannot be constant as a result of expansion or contraction in the gravitational field.

To define the string scalars $\theta$, $\kappa$ one sets  at each point of the string trajectory a tetrade $l,n,p,q$. 
Here $p=\eta/|\eta|$, $|\eta|^2=(\eta \cdot \eta)$, $n$ is null, orthogonal to $p$ and normalized as
$(n\cdot l)=-2$, vector $q$ is spacelike, unit, and orthogonal to $l,n,p$. 
Condition $|\eta|\neq 0$ is assumed.

The parameters $\theta$ and $\kappa$ are introduced as the following spin coefficients \cite{Fursaev:2021xlm} :
\begin{equation}\label{1.4}
\theta=(p \cdot \nabla_p l)~~~,~~~ \kappa=(q \cdot \nabla_p l)~~.
\end{equation} 
(Relation to notations of \cite{Fursaev:2021xlm} is $\theta=\theta_s$, $\kappa=\kappa_2$).
Spin coefficients (\ref{1.4}) are invariant with respect to null rotations of the tetrade: 
\begin{equation}\label{1.6}
l'=l~~,~~p'=p~~,~~n=n'+2\omega q'+\omega^2 l'~~, ~~q=q'+\omega l'~~~.
\end{equation} 
Rotations (\ref{1.6}) and reparametrizations of $\lambda,\tau$,
\begin{equation}\label{1.9}
\lambda'=g(\lambda,\tau)~~,~~\tau'=\phi(\tau)~~,
\end{equation} 
make a 2-parameter group of $l$-preserving null rotations of $n,p,q$, accompanied with rescalings of
$l$ and $n$, see \cite{Fursaev:2021xlm}. Parameters $\theta$ and $\kappa$ transform 
as boost-weighted scalars, $Q=(g_{,\lambda})^bQ'$, with boost weight $b=1$.

We also use a pair of null complex vectors: 
\begin{equation}\label{1.6a}
m={1 \over \sqrt{2}}(p+iq)~~,~~\bar{m}={1 \over \sqrt{2}}(p-iq)~~.
\end{equation}
Invariants in optical equation (\ref{i.1}) are
\begin{equation}\label{1.6b}
\Psi_0=-C_{mlml}~~,~~\Phi_{00}=-\frac 12 R_{ll}~~,
\end{equation}
where $C_{abcd}$ and $R_{ab}$ are components, respectively, of the Weyl tensor Ricci tensor in the 
given basis, see \cite{Penrose:1986ca}, \cite{Chandrasekhar:1985kt}. Definitions (\ref{1.6b}) are introduced on the string trajectory.

\subsection{Stress-energy tensor and energy conservation}\label{energy}

We start with the case of a straight string in Minkowsky spacetime. If the string is parallel to the $z$-axis and moves along the $x$-axis its stress-energy tensor is
\begin{equation}\label{1.5}
T^{\mu\nu}(x)=\mu \delta(y)\delta(u) l^\mu l^\nu~~,
\end{equation} 
where $u=t-x$, $l^\mu\partial_\mu=\partial_t+\partial_x$. The parameter $\mu$ is the energy of the string per unit
length. Stress-tensor (\ref{1.5}) was found in \cite{Fursaev:2017aap} by applying the Aichelburg-Sexl boost
\cite{Barrabes:2002hn} to the stress-energy tensor of a straight massive (tensile) cosmic string. 
The same boost applied to the Riemann tensor results in a non-trivial component $R_{uyuy}=\omega \delta(y)\delta(u)$, where 
$\omega=8\pi G\mu$. The null string leaves the spacetime locally flat but creates a non-trivial holonomy
at the string worldsheet: a parallel transport 
of a vector around a point of the string trajectory results in a null rotation defined by (\ref{1.6}) 
with $\omega=8\pi G\mu$.

As it is easy to see, definition (\ref{1.5}) implies a standard covariant conservation law
\begin{equation}\label{1.7}
\nabla_\mu T^{\mu\nu}(x)=0~~,
\end{equation} 
for a constant parameter $\mu$, or, in general, if 
\begin{equation}\label{1.8}
\left. \partial_l\mu \right|_{u=y=0}=0~~.
\end{equation}
The last condition allows $\mu$ to vary along the string, $\mu=\mu(z)$. Such a string creates a locally flat spacetime with a non-constant holonomy, where $\omega=\omega(z)=8\pi G\mu(z)$.

Consider now a null string in an arbitrary spacetime $\cal M$ with the trajectory 
$x^\mu=\bar{x}^\mu(\lambda,\tau)$. A natural generalization of (\ref{1.5}) is the following stress-energy tensor of the string:
\begin{equation}\label{1.10}
T^{\mu\nu}(x)=\int d\lambda d\tau~ \bar{\mu}(\lambda,\tau)~ \delta^{4}(x,\bar{x})~ l^\mu(\bar{x}) l^\nu(\bar{x})~~,
\end{equation} 
where $\bar{\mu}(\lambda,\tau)$ is some density, $\delta^{4}(x,\bar{x})=\delta^{4}(x-\bar{x})/\sqrt{-g}$ is an
invariant delta-function with the support on the string trajectory, $l^\mu(\bar{x})=d\bar{x}^\mu/d\lambda$. 
Note that $\bar{\mu}(\lambda,\tau)$
transforms as
\begin{equation}\label{1.11}
\bar{\mu}(\lambda,\tau)=\phi_{,\tau} g^{-1}_{,\lambda}~\bar{\mu}'(\lambda',\tau')~~
\end{equation} 
under reparametrizations (\ref{1.9}).
After some algebra one can check that covariant conservation law (\ref{1.7}) holds for SET (\ref{1.10}) 
under the condition:
\begin{equation}\label{1.12}
\partial_\lambda \bar{\mu}+\beta \bar{\mu}=0~~,
\end{equation} 
where $\beta$ is defined in (\ref{1.3}). In what follows we assume that $\lambda$ is affine parameter. In 
this case $\bar{\mu}$ does not depend on $\lambda$.

It is instructive to see how definition (\ref{1.10}) reduces to (\ref{1.5}) in case of straight string in Minkowsky spacetime. String trajectory is given by equations: $t=\lambda$, $x=\lambda$, $z=\tau$, $y=0$.
Integration over $\lambda$ in (\ref{1.10}) yields $\delta(u)$, integration over $\tau$ results in (\ref{1.5})
with $\mu=\bar{\mu}(z)$.

One expects a null string creates a local holonomy around each point $x^\mu=\bar{x}^\mu(\lambda,\tau)$ with a parameter 
$\omega=\omega(\lambda,\tau)$. By using example of a straight string in Minkowsky
spacetime,  the {\it physical energy} of the string $\mu=\mu(\lambda,\tau)$ can be introduced  
as a parameter which determines  the local holonomy by relation $\omega(\lambda,\tau)=8\pi G\mu(\lambda,\tau)$. 
To understand connection between $\bar{\mu}$ and $\mu$ we need to
bring SET (\ref{1.10}), at a chosen point 
$x^\mu_\ast=\bar{x}^\mu(\lambda_\ast,\tau_\ast)$, to ``flat form'' (\ref{1.5}). 
This can be done in local coordinates, where
$g_{\mu\nu}(x_\ast)=\eta_{\mu\nu}$, $\Gamma^\rho_{\mu\nu}(x_\ast)=0$, $x_\ast=0$, and 
string equations (\ref{1.1})-(\ref{1.3}) have a simple solution near $x_\ast$, 
\begin{equation}\label{1.13}
t \simeq \lambda-\lambda_\ast~~,~~x\simeq \lambda-\lambda_\ast~~,~~y=0~~,~~z\simeq z_{,\tau}(\tau_\ast)
(\tau-\tau_\ast)~~.
\end{equation} 
Then $T^{\mu\nu}(x)$ near $x_\ast$ takes form (\ref{1.5}) with the physical energy
\begin{equation}\label{1.14}
\mu(\lambda,\tau)=\bar{\mu}(\tau)|\eta(\lambda,\tau)|^{-1}~~,
\end{equation} 
where factor $|\eta(\lambda,\tau)|=|z_{,\tau}(\tau_\ast)|$ appears when integrating over $\tau$.
The key difference between $\mu$ and $\bar{\mu}$ is that $\mu$ does not depend on reparametrizations
of $\tau$, see (\ref{1.9}). Stress-energy tensor (\ref{1.10}) in terms of the physical energy looks as
\begin{equation}\label{1.15}
T^{\mu\nu}(x)=\int d\lambda ~|\eta| d\tau~ \mu~ \delta^{4}(x,\bar{x})~ l^\mu(\bar{x}) l^\nu(\bar{x})~~.
\end{equation} 
Element $|\eta|d\tau$ is the physical length of a segment of the string between $\bar{x}^\mu(\lambda,\tau)$
and $\bar{x}^\mu(\lambda,\tau+d\tau)$.

Energy conservation law (\ref{i.2}) follows from (\ref{1.14}) and definition (\ref{1.4}) of $\theta$.

\subsection{Caustics of null strings}\label{caus}

Since the string energy $\mu(\lambda,\tau)$ is determined with respect to the physical length of a string segment, it 
develops singularities at points where connecting vector $\eta$ has vanishing
norm, $\eta^2=0$. These may be isolated points or a one-parameter family $\lambda=\lambda_c(\tau)$ 
on the string trajectory which makes
a curve
\begin{equation}\label{1.16}
x^\mu_c(\tau)\equiv x^\mu(\lambda_c(\tau),\tau)~~.
\end{equation} 
Condition $\eta^2=0$ implies that either $\eta=0$ or $\eta$ is a null vector at $\lambda=\lambda_c(\tau)$.
If $\eta=0$, the tangent vector $\zeta$ to curve (\ref{1.16}) is $\zeta=\dot{x}_c=\dot{\lambda}_c~l$. That is,
$\zeta$ is null and directed along $l$.
If $\eta$ becomes null at $\lambda=\lambda_c(\tau)$, it follows from (\ref{1.2}) that $\eta$ should be directed along $l$.
Then $\zeta=\dot{\lambda}_c~l+\eta$ is null and it is again directed 
along $l$. Therefore,  curve (\ref{1.16}) is a light-like caustic where trajectories of different points of the string are tangent to. 

Optical equation (\ref{i.1}) implies the following behavior of the string scalar and string energy near a caustic:
\begin{equation}\label{1.17}
Z(\lambda,\tau)\sim {1 \over \lambda -\lambda_c(\tau)}~~,
\end{equation} 
\begin{equation}\label{1.18}
\mu (\lambda,\tau)\sim {\mu_c(\tau) \over |\lambda -\lambda_c(\tau)|}~~,
\end{equation} 
where $\mu_c(\tau)$ is some function. 

Caustics of null strings, like caustics in optics,  are the regions where energy is concentrated. For 
this reason caustics are distinctive properties of null strings which may result in important physical effects.

\subsection{Noether charges of null strings}\label{char}

By using SET (\ref{1.10})
one can define conserved charges $Q$ of the null string. If a spacetime admits a Killing vector field $\zeta$, 
\begin{equation}\label{2.19}
Q(\zeta)=\int_{\Sigma}d\Sigma_\mu ~J^\mu(\zeta)~~,~~J^\mu(\zeta)=T^{\mu\nu}\zeta_\nu~~.
\end{equation}
Integration in (\ref{2.19}) goes over a Cauchy hypersurface $\Sigma$.

We consider null strings in asymptotically flat  spacetimes and use (\ref{2.19}) to define ``in'' and ``out'' linear and angular 
momenta 
of the string at past null infinity $\mathcal{I}^{-}$  and future null infinity $\mathcal{I}^{+}$.  Points of null strings 
near $\mathcal{I}^{\pm}$ move as almost radial light rays.
The metric near $\mathcal{I}^{-}$, $\lambda\to -\infty$, can be written as
\begin{equation}\label{2.20}
ds^2\simeq -dv^2+2dvdr+r^2d\Omega^2~~. 
\end{equation}
The string trajectory is $r_{\mathrm{in}}\simeq|\lambda|$, $v_{\mathrm{in}}=0$,
$\theta_{\mathrm{in}}=\theta_{\mathrm{in}}(\tau)$, $\varphi_{\mathrm{in}}=\varphi_{\mathrm{in}}(\tau)$.
Near $\mathcal{I}^{+}$, $\lambda\to \infty$,
\begin{equation}\label{2.21}
ds^2\simeq -du^2-2dudr+r^2d\Omega^2~~, 
\end{equation}
$r_{\mathrm{out}}\simeq \lambda$, $u_{\mathrm{out}}=0$,
$\theta_{\mathrm{out}}=\theta_{\mathrm{out}}(\tau)$, $\varphi_{\mathrm{out}}=\varphi_{\mathrm{out}}(\tau)$. The 
corresponding conserved charges near $\mathcal{I}^{-}$ and 
$\mathcal{I}^{+}$, respectively, are
\begin{equation}\label{2.22a}
Q_{\mathrm{in}}(\zeta)=\int_{r=R} \sqrt{-g} J_{\mathrm{in}}^r(\zeta)dv~d\theta d\varphi~~,~~Q_{\mathrm{out}}(\zeta)=\int_{r=R} \sqrt{-g} J_{\mathrm{out}}^r(\zeta)du~d\theta d\varphi~~.
\end{equation}
Surfaces $\Sigma_{\mathrm{in}}$ and $\Sigma_{\mathrm{out}}$ in (\ref{2.22a}) are taken at $r=R$.
One finds with the help of  (\ref{1.10}), (\ref{2.19}) 
\begin{equation}\label{2.22}
Q_{\mathrm{in}}(\zeta)=\int d\tau \bar{\mu}(\tau)(l_{\mathrm{in}}\cdot \zeta)~~,~~Q_{\mathrm{out}}(\zeta)=\int d\tau \bar{\mu}(\tau)(l_{\mathrm{out}}\cdot \zeta)~~.
\end{equation}
The integration in (\ref{2.22a}) over $u,v,\theta, \varphi$ is performed by using equations of radial trajectories.

\section{Scattering of null strings by black holes}\label{scatt}
\setcounter{equation}0
\subsection{Strings in Minkowsky spacetime}\label{flat}

To give preliminary definitions we begin with null strings in Minkowsky spacetime. 
If $\lambda$ is an affine parameter, general solution to (\ref{1.1}) -- (\ref{1.3}) 
is
\begin{equation}\label{2.1}
x^\mu(\lambda,\tau)=\lambda b^\mu(\tau)+a^\mu(\tau)~~,
\end{equation}
where $b^\mu$ is an arbitrary null vector, $b^2=0$. Restrictions on $a^\mu$ are: $(b\cdot \dot{a})=0$,
see (\ref{1.2}), and $\dot{a}^2>0$, $\dot{a}\equiv a_{,\tau}$.  One finds:
\begin{equation}\label{2.2}
|\eta|^2=\lambda^2 \dot{b}^2+2\lambda (\dot{b} \cdot \dot{a})+\dot{a}^2~~.
\end{equation}

A residual freedom, $\lambda \to\lambda+g(\tau)$, can be fixed by additional 
physical conditions, for example, 
$(l\cdot u_o)=-1$, $(\eta \cdot u_o)=0$, where  $u_o$ is the 4-velocity of observers, 
see \cite{Fursaev:2021xlm}. In Minkowsky coordinates, where velocity of observers is $u_o^\mu=\delta^\mu_0$, these
conditions are ensured if $t(\lambda,\tau)=\lambda$, that is $\dot{a}$ and $\dot{b}$ have only spatial components.
Let $|\dot{b}|\neq 0$, then \cite{Fursaev:2021xlm}
\begin{equation}\label{2.4}
Z(\lambda,\tau)={1 \over \lambda+z_0(\tau)}~~,~~z_0(\tau)=r(\tau) e^{i\varphi (\tau)}~~.
\end{equation}
\begin{equation}\label{2.3}
\cos\varphi\equiv {(\dot{b} \cdot \dot{a}) \over |\dot{a}||\dot{b}|}~~,~~r\equiv { |\dot{a}| \over |\dot{b}|}~~,
\end{equation}
$Z(\lambda,\tau)$ is a solution to (\ref{i.1}) with $\Psi_0=\Phi_{00}=0$. 

It is easy to see from (\ref{2.2}) that caustics of null strings discussed in Sec.\ref{caus} appear when $\dot{a}(\tau)=-\lambda_c(\tau) \dot{b}(\tau)$, $\eta=0$.  In this case $Z(\lambda,\tau)=1/(\lambda-\lambda_c(\tau))$, in agreement with (\ref{1.17}). 

One can represent $Z(\lambda,\tau)$ in another form. For example, if $\bar{\theta}(\tau)$ and $\bar{\kappa}(\tau)$
are expansion and rotation of the string, say, at $\lambda=0$, then
\begin{equation}\label{2.6}
Z(\lambda,\tau)={\bar{\theta}(\tau)+i\bar{\kappa}(\tau) \over \lambda(\bar{\theta}(\tau)+i\bar{\kappa}(\tau))+1 }
\end{equation}
Caustics in (\ref{2.6}) appear when $\bar{\kappa}(\tau)=0$, 
$\lambda=\lambda_c(\tau)=-1/\bar{\theta}(\tau)$. 

As for $\bar{\theta}(\tau)$ and $\bar{\kappa}(\tau)$, these parameters can be expressed in terms of more convenient
characteristics of a ``snapshot'' of the string at $\lambda=0$. One can use the Frenet frame,
\begin{equation}\label{2.7}
\vec{E}_1 = \frac{\dot{\vec{a}}}{|\dot{\vec{a}}|}~~,~~\dot{\vec{E}}_1 = k_1 \vec{E}_2~~,~~
\dot{\vec{E}}_2 = -k_1 \vec{E}_1 + k_2 \vec{E}_3~~,~~
\dot{\vec{E}}_3 = -k_2 \vec{E}_2
\end{equation}
to define curvature $k_1(\tau)$ and torsion $k_2(\tau)$ of curve $\vec{x}(0,\tau)=\vec{a}(\tau)$. Vector $\vec{b}$ can be chosen as
\begin{equation}\label{2.8}
\vec{b} (\tau) = \cos\alpha (\tau)~\vec{E}_2 + \sin\alpha (\tau)~ \vec{E}_3~~.
\end{equation}
It can be shown that
\begin{equation}\label{2.9}
\bar{\theta}= - \frac{ k_1 \cos\alpha}{ |\dot{\vec{a}}| }~~,~~\bar{\kappa}=\frac{ \dot{\alpha} +  k_2 }
{ |\dot{\vec{a}}|}
\end{equation}
Thus, the curvature of $\vec{x}(0,\tau)$ determines expansion of the string, while its rotation depends on the torsion of $\vec{x}(0,\tau)$.

There is a particular class of strings with $\dot{b}\equiv 0$. For such a class $Z\equiv 0$.
From (\ref{2.6}) one concludes that in this case $\bar{\theta}(\tau)=\bar{\kappa}(\tau)=0$, and $\dot{\alpha}=0$, see (\ref{2.8}).
Equations (\ref{2.9}) imply that $k_2=0$, and $k_1=0$ or $\cos\alpha=0$. This means that the string is a straight line, 
or the curve $\vec{x}(0,\tau)=\vec{a}(\tau)$ lies in a plane, and $\vec{b}$ is orthogonal to the plane, $(b\cdot \dot{a})=
(b\cdot \ddot{a})=0$.

We call strings with $Z\equiv 0$ frozen strings, since they preserve their form during the evolution.
A simple example of the frozen string is a straight string.

\subsection{Scattering by non-rotating black holes}\label{weak1}

The aim of this Section is to study scattering of a null string by a black hole, or a gravitating source, in general.  Analysis of
string equations (\ref{1.1}) - (\ref{1.3}) for null strings 
near massive sources can be quite complicated \cite{Kar:1995br}-\cite{Dabrowski:2002iw}. We do
calculations in the weak field approximation for strings which move far from the sources, that is, with large impact parameters. 
Our aim is to understand some universal features of the string dynamics. Analogous scattering problems for tensile strings
have been studied in  \cite{Lonsdale:1988x},\cite{DeVilliers:1998xz},\cite{Page:1998ya}. Scattering on non-rotating black holes
can be found in \cite{DeVilliers:1998nm}

Suppose a black hole is located at the center of coordinates $x=y=z=0$. 
The mass of the black hole and its angular momentum are, respectively,  $M$, $J^i$. 
In the weak-field approximation the flat metric 
\begin{equation}\label{2.10}
ds^2=-dt^2+dx^2+dy^2+dz^2~~
\end{equation}
acquires corrections
\begin{equation}\label{2.10a}
\delta g_{\mu\nu}=h^M_{\mu\nu}+h^J_{\mu\nu}~~
\end{equation}
with the following non-zero components:
\begin{equation}\label{2.11a}
h^M_{00}=\frac{r_g}{r}~~,~~h^M_{ij}=\frac{r_g\,x_i x_j}{r^3}~~,
\end{equation}
\begin{equation}\label{2.11b}
h^J_{0i}=2G~\varepsilon_{ijk}\frac{x^j J^k}{r^3}~~,
\end{equation}
where $r_g=2MG$, $r=\sqrt{x^2+y^2+z^2}$.  We denote coordinates of the string trajectory  at 
$\mathcal{I}^{-}$ by $x_{\mathrm{in}}$, 
\begin{equation}\label{2.12a}
x(\lambda,\tau)_{\lambda\to-\infty}\to x_{\mathrm{in}}(\lambda,\tau)=\lambda b_{\mathrm{in}}(\tau)+
a_{\mathrm{in}}(\tau)~~,
\end{equation}
and by $x_{\mathrm{out}}$ at  $\mathcal{I}^{+}$,
after the scattering on the source,
\begin{equation}\label{2.12b}
x(\lambda,\tau)_{\lambda\to \infty}\to
x_{\mathrm{out}}(\lambda,\tau)=\lambda b_{\mathrm{out}}(\tau)+
a_{\mathrm{out}}(\tau)~~.
\end{equation}
Velocities of the string are $l_{\mathrm{in}}=b_{\mathrm{in}}$, at $\mathcal{I}^{-}$, and
$l_{\mathrm{out}}=b_{\mathrm{out}}$, at $\mathcal{I}^{+}$.

\begin{figure}[h]
\begin{center}
\includegraphics[height=12cm,width=14cm]{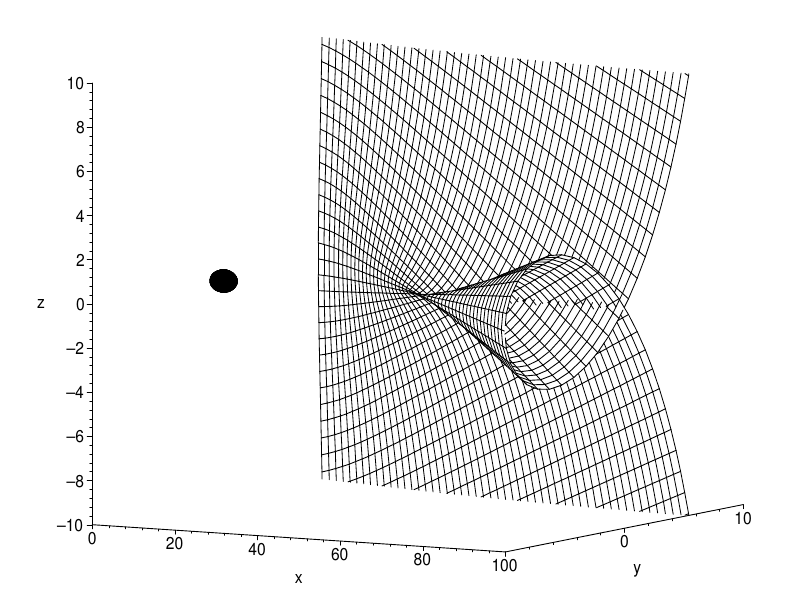}
%\v_space{0.4cm}
\caption{\small{shows the world-sheet of a straight cosmic string moving near a non-rotating black hole
located at the center of coordinates. The ratio of the gravitational radius to the impact parameter is $0.1$.
The grid is made of lines $\lambda=\mathrm{const}$, $\tau=\mathrm{const}$.}}
\label{Fig_WS}
\end{center}
\end{figure}

For further use it is convenient to introduce a minimal distance $r_0(\tau)$ between the center of coordinates
and a segment of the string with parameter $\tau$. In the absence of the source, $r^2=\lambda^2+
2\lambda (b_{\mathrm{in}}\cdot a_{\mathrm{in}})+a_{\mathrm{in}}^2$ has a minimum 
at $\lambda=\lambda_0=-(b_{\mathrm{in}}\cdot a_{\mathrm{in}})$. One can define
\begin{equation}\label{2.13}
\vec{x}_0(\tau)=\vec{x}_{\mathrm{in}}(\lambda_0,\tau)=\vec{a}_{\mathrm{in}}-\vec{b}_{\mathrm{in}}
(b_{\mathrm{in}}\cdot a_{\mathrm{in}})~~,
~~r_0(\tau)=|\vec{x}_0(\tau)|=\sqrt{a^2_{\mathrm{in}}-
(b_{\mathrm{in}}\cdot a_{\mathrm{in}})^2}~~.
\end{equation}

We are looking for effects of metric perturbations (\ref{2.11a}), (\ref{2.11b}) 
in the linear approximation in $M$ and $J$, $x=x_{\mathrm{in}}+\delta_M x+\delta_J x$, when change of the string trajectory
is determined by linearized version of (\ref{1.3}) (with $\beta=0$)
\begin{equation}\label{2.14}
\partial_\lambda^2 ~\delta x^{\mu}+\delta\Gamma^{\mu}_{\alpha\beta}(x_{\mathrm{in}})~ l_{\mathrm{in}}^\alpha 
l_{\mathrm{in}}^\beta =0~~.
\end{equation}
Here $\delta\Gamma^{\mu}_{\alpha\beta}(x_{\mathrm{in}})$ are calculated on  
$x_{\mathrm{in}}$ with the help of (\ref{2.11a}), (\ref{2.11b}). This is the scattering with large impact parameter, $r_0(\tau)\gg r_g$.

In this Section we assume that the black hole is non-rotating,
$J^i=0$. At large $r$ the effects of the non-zero angular 
momentum are weaker than effects caused by the mass. They are considered in Sec. \ref{angm}.

\begin{figure}[h]
\begin{center}
\includegraphics[height=8cm,width=13cm]{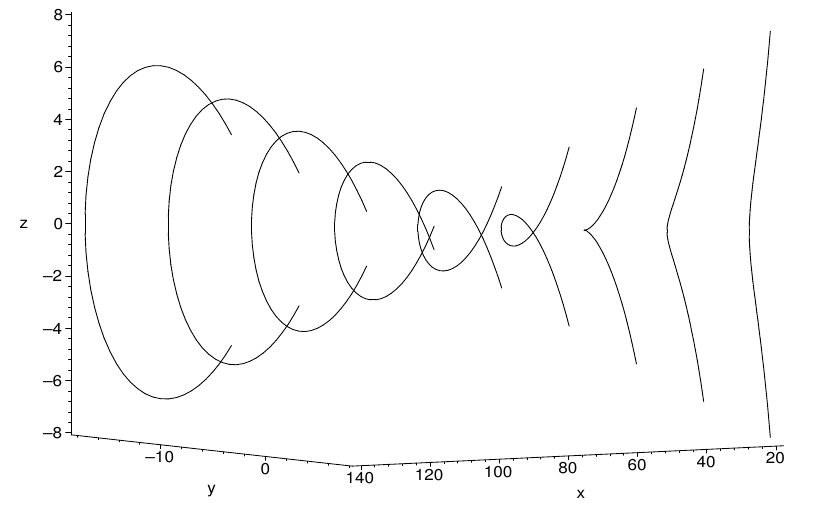}
%\v_space{0.4cm}
\caption{\small{depicts scattering of a finite segment of a straight cosmic string 
by a non-rotating black hole. The parameters are the same as for Fig. \ref{Fig_WS}.  Before the string
forms a loop a point-like caustic appears 
at the distance of 5 impact parameters behind the black hole.}}
\label{Fig_String_evolution}
\end{center}
\end{figure}

It is clear that solutions to (\ref{2.14}) differ by additions of the form $\lambda c^\mu(\tau)+d^\mu(\tau)$,
where $c^\mu(\tau)$ and $d^\mu(\tau)$ are arbitrary 4-vectors. The solution is uniquely fixed by requiring 
conditions (\ref{1.1}), (\ref{1.2}), and $l^2=0$. We also require that $(l\cdot u_o)\to -1$, 
$(\eta\cdot u_o)\to 0$ at $\mathcal{I}^{+}$, where $u_o^\mu=\delta^\mu_0$ is 4-velocity of the chosen set 
of observers.  

The solution to (\ref{2.13}) which satisfies these conditions in the linear approximation 
in $r_g$ is
\begin{equation}\label{2.15a}
t(\lambda,\tau)=\lambda-r_g~\ln \left[{r_{\mathrm{in}}-\lambda-(a_{\mathrm{in}}\cdot b_{\mathrm{in}}) \over \varrho}\right]~~,
\end{equation}
\begin{equation}\label{2.15b}
\vec{x}(\lambda,\tau) = \vec{x}_{\mathrm{in}}(\lambda,\tau) -r_g~{\vec{x}_0(\tau)  \over r_{\mathrm{in}}-\lambda-(a_{\mathrm{in}}\cdot b_{\mathrm{in}})}
~~,
\end{equation}
where $\varrho$ is some dimensional parameter.

\begin{figure}[h]
\begin{center}
\includegraphics[height=8cm,width=12cm]{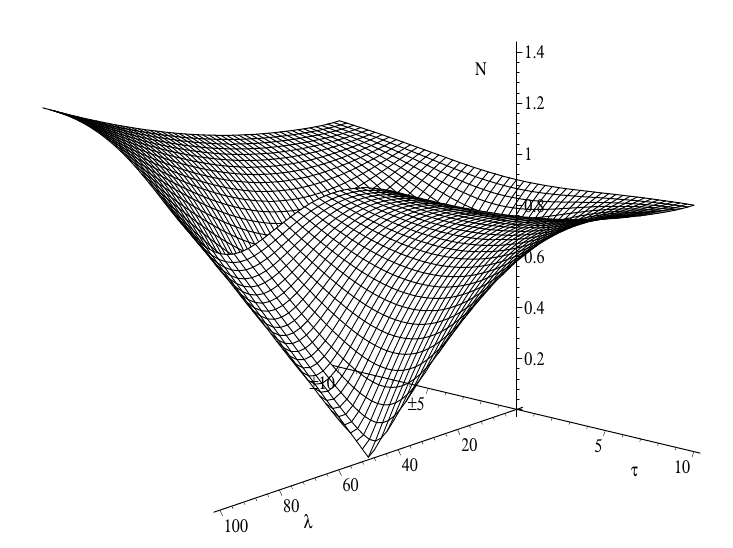}
%\v_space{0.4cm}
\caption{\small{shows norm $|\eta|$ of the connecting vector of the straight string during the  scattering on the black
hole. The parameters are the same as for Figs. \ref{Fig_WS}, \ref{Fig_String_evolution}. The caustic forms at a point where
$|\eta|=0$.}}
\label{Fig_N}
\end{center}
\end{figure}

We need scattering data at $\mathcal{I}^{+}$, where (\ref{2.15a}), (\ref{2.15b}) become
\begin{equation}\label{2.16a}
t(\lambda,\tau)\simeq \lambda-r_g~\ln {r_0^2 \over \lambda \varrho}~~,
\end{equation}
\begin{equation}\label{2.16b}
\vec{x}(\lambda,\tau)\simeq \lambda\vec{b}_{\mathrm{out}}+\vec{a}_{\mathrm{out}}
=\vec{x}_{\mathrm{out}}(\lambda,\tau)
~~,
\end{equation}
\begin{equation}\label{2.16c}
\vec{b}_{\mathrm{out}}=\vec{b}_{\mathrm{in}}-\Phi~\vec{n}~~,~~~~\Phi={2r_g \over r_0}~~,
\end{equation}
\begin{equation}\label{2.16d}
\vec{a}_{\mathrm{out}}=\vec{a}_{\mathrm{in}}-\Phi~ (a_{\mathrm{in}}\cdot b_{\mathrm{in}})~\vec{n}
~~,~~\vec{n}={\vec{x}_0 \over r_0}~~,
\end{equation}
One can check that $|\vec{b}_{\mathrm{out}}|=1$ up to terms $O(\Phi^2)$.
Thus, scattering results in rotation of the string velocity 
$l_{\mathrm{in}}=b_{\mathrm{in}}$ in the plane $b_{\mathrm{in}},a_{\mathrm{in}}$ by angle $\Phi$,
\begin{equation}\label{2.17}
l^i_{\mathrm{out}}=M_{ij} ~l^j_{\mathrm{in}}~~,
\end{equation}
\begin{equation}\label{2.18}
M_{ij}\simeq \delta_{ij}+\Phi~\omega_{ij}~~,~~\omega_{ij}=r_0^{-1}(b_{\mathrm{in}}^i a_{\mathrm{in}}^j-b_{\mathrm{in}}^j 
a_{\mathrm{in}}^i)~~.
\end{equation}
This effect is similar to deflection of light rays by massive bodies in general relativity.

It should be pointed out that even in the weak field approximation evolution of a null string described by (\ref{2.15a}), (\ref{2.15b}) may 
be non-trivial due to creation of loops and caustics. The simplest illustration how null strings form loops 
is the scattering of a straight string. The world-sheet of the string is shown on Figure \ref{Fig_WS}.  The  string equation 
at  $\mathcal{I}^{-}$ is taken as
$t=x=\lambda$, $z=\tau$, $y=\rho$, where $\rho$ is an impact parameter. The black hole is located at the center of coordinates.  
The results are obtained for the ratio $2MG/\rho=0.1$.
Figure \ref{Fig_String_evolution}  
demonstrates scattering of a finite segment of the same string and creation of a caustic at an isolated point.
Figure \ref{Fig_N} depicts the norm $|\eta|$ of the connecting vector. The caustic appears when $|\eta|=0$. 

It is instructive to compare these results with scattering of tensile cosmic strings \cite{Lonsdale:1988x},
\cite{DeVilliers:1998xz}, \cite{Page:1998ya}, \cite{DeVilliers:1998nm}.
Astrophysical tensile cosmic strings are expected to move with the velocity $v\simeq 0,7 c$. At such velocities straight tensile strings, as a result
of scattering by a black hole, are displaced in the direction perpendicular to their motion \cite{DeVilliers:1998nm}.
At velocities $0,9 c<v<1 c$ scattering of ultrarelativistic tensile cosmic strings reveals formation of loops
which is similar to the case of null strings \cite{DeVilliers:1998nm}. As has been 
pointed out in \cite{Page:1998ya}, in the ultrarelativistic limit
of a string moving very near the speed of light in a direction perpendicular to the
string, the propagation of disturbances along the string suffers a large time dilation,
so that each piece of the string is effectively decoupled and moves very nearly along
a null geodesic.

As a next step, we use Sec. \ref{char} to analyze how energy, linear momentum and angular momentum of the string
are changed during the scattering.
The energy
corresponds to the Killing vector $\zeta=\partial_t$ and it conserves,
\begin{equation}\label{2.23}
E_{\mathrm{in}}=\int d\tau \bar{\mu}(\tau)=E_{\mathrm{out}}~~.
\end{equation}
For linear momenta ($\zeta=\partial_i$) one has
\begin{equation}\label{2.24}
\vec{P}_{\mathrm{in}}=\int d\tau \bar{\mu}(\tau)\vec{l}_{\mathrm{in}}~~,~~
\vec{P}_{\mathrm{out}}=\int d\tau \bar{\mu}(\tau)\vec{l}_{\mathrm{out}}~~.
\end{equation}
\begin{equation}\label{2.25}
\vec{P}_{\mathrm{out}}=\vec{P}_{\mathrm{in}}-\int d\tau \bar{\mu}(\tau)\Phi(\tau)~\vec{n}(\tau)~~.
\end{equation}

Consider an axis which goes through the center of coordinates and is directed along a unit vector $\vec{Z}$.
The angular momentum $j(Z)$  of the string related to rotations around the given axis
is a charge for the Killing field $\zeta^i=-\varepsilon_{ijk}x^j Z^k$. The angular momentum 
at $\mathcal{I}^{-}$  is
\begin{equation}\label{2.26}
j_{\mathrm{in}}(Z)=-\int d\tau \bar{\mu}(\tau)(\vec{l}_{\mathrm{in}}\cdot 
[\vec{x}_{\mathrm{in}} \times \vec{Z}])=-\int d\tau \bar{\mu}(\tau)(\vec{l}_{\mathrm{in}}\cdot 
[\vec{a}_{\mathrm{in}} \times \vec{Z}])~~.
\end{equation}
Change of the angular momentum at $\mathcal{I}^{+}$ depends on variations of $l_{\mathrm{in}}$ and 
$a_{\mathrm{in}}$,
\begin{equation}\label{2.27}
\delta j(Z)=-\int d\tau \bar{\mu}(\tau)\left((\delta \vec{l}\cdot 
[\vec{a}_{\mathrm{in}} \times \vec{Z}])+(\vec{l}_{\mathrm{in}}\cdot 
[\delta\vec{a} \times \vec{Z}])\right)~~.
\end{equation}
However, if (\ref{2.16c}) and (\ref{2.16d}) are used in (\ref{2.27}), the variation vanishes,
\begin{equation}\label{2.28}
\delta j(Z)=-\int d\tau \bar{\mu}(\tau)~\Phi~(\vec{x}_0\cdot 
[\vec{n} \times \vec{Z}])=0~~.
\end{equation}
The string does not change its angular momentum.

By the conservation laws the black hole after interaction with the string receives 
a non-zero linear momentum 
\begin{equation}\label{2.29}
\vec{P}_{\mathrm{body}}=\vec{P}_{\mathrm{in}}-\vec{P}_{\mathrm{out}}=\int d\tau \bar{\mu}(\tau)\Phi(\tau)~\vec{n}(\tau)~~,
\end{equation}
see (\ref{2.25}).This is the gravitational recoil effect. 

To give an example of the recoil effect consider again the scattering of the straight string. Suppose that
string energy $\mu$ is constant. Calculation of the recoil momentum yields 
the only non-zero component, along $y$ axis,
\begin{equation}\label{2.30a}
P^y_{\mathrm{body}}=2\pi r_g \mu~~.
\end{equation}
The corresponding change of the energy of the body, $\delta E\sim MG^2\mu^2$, can be neglected
in the linear approximation. The momentum may be different if
string energy depends on $z$. For example, another component of the momentum, $P^z_{\mathrm{body}}$, may be non-trivial.

\subsection{Scattering by rotating black holes}\label{angm}

We focus now on physical effects related to angular momenta of black holes. 
Scattering of tensile cosmic strings on rotating black holes
can be found in \cite{Snajdr:2002rw}-\cite{Dubath:2006vs}. Angular momentum transfer from rotating black holes to tensile strings may result in interesting effects. For example,
a tensile cosmic string piercing a rotating black hole may spin-down the black hole \cite{Boos:2017pyd} such that the angular momentum vector
of the black hole aligns with the string \cite{Xing:2020ecz}.

To proceed we put $h_{\mu\nu}^M=0$
in (\ref{2.10a}).
The solution  to (\ref{2.14}) with the same choice of asymptotic conditions which 
lead to (\ref{2.15a}), (\ref{2.15b}) is the following:
\begin{equation}\label{j2.15a}
t(\lambda,\tau)=\lambda-{2G~r_0 \over  r_{\mathrm{in}}(r_{\mathrm{in}}-\lambda-(a_{\mathrm{in}}\cdot b_{\mathrm{in}}))}(J\cdot c)~~,
\end{equation}
\begin{equation}\label{j2.15b}
\vec{x}(\lambda,\tau) = \vec{x}_{\mathrm{in}}(\lambda,\tau) +2G~\frac{2(J\cdot c)~
[\vec{b}_{\mathrm{in}}\times\vec{c}]+[\vec{J}\times\vec{d}]}
{r_{\mathrm{in}}-\lambda-(a_{\mathrm{in}}\cdot b_{\mathrm{in}})}-
2G~\frac{(J\cdot c)}{r_0} (\vec{d}+\vec{b}_{\mathrm{in}})
~~,
\end{equation}
where 
\begin{equation}
\vec{c}=\frac{[\vec{a}_{\mathrm{in}}\times\vec{b}_{\mathrm{in}}]}{r_0}~~,~~ 
\vec{d}=\frac{\lambda~ \vec{b}_{\mathrm{in}}+\vec{a}_{\mathrm{in}}}{r_{\mathrm{in}}}~~.
\end{equation}
With the help of (\ref{j2.15a}), (\ref{j2.15b}) one finds that the angular momentum results in 
rotation of initial vectors $a_{\mathrm{in}}$, $b_{\mathrm{in}}$ which determine trajectory of the string at 
$\mathcal{I}^{-}$,
\begin{equation}\label{j2.16c}
b^i_{\mathrm{out}}=b^i_{\mathrm{in}}+\frac{4G}{r_0^2}\varepsilon_{ijk}~b^j_{\mathrm{in}} \left(2(J\cdot c)~c^k-J^k\right)~~,
\end{equation}
\begin{equation}\label{j2.16d}
a^i_{\mathrm{out}}=a^i_{\mathrm{in}}+\frac{4G(J\cdot b_{\mathrm{in}})}{r_0^2}\varepsilon_{ijk}~a^j_{\mathrm{in}}b^k_{\mathrm{in}}+\frac{4G(J\cdot c)}{r_0^2}\left(2(a_{\mathrm{in}}\cdot b_{\mathrm{in}}) n^i-r_0~ b^i_{\mathrm{in}}\right)~~.
\end{equation}
Variation of the angular momentum of the string can be computed with the help of (\ref{2.22}), (\ref{j2.16c}),  (\ref{j2.16d}),
\begin{equation}\label{j2.27}
 \delta j(Z)=4G\int d\tau\frac{\bar\mu}{r_0^2} (Z\cdot [\vec{J}\times[\vec{a_{\mathrm{in}}}\times\vec{b_{\mathrm{in}}}]])~~.
\end{equation}
Equation (\ref{j2.27}) can be interpreted 
as a  spin-spin interaction. Variation of the angular momentum due to the mass of the source vanishes, see  (\ref{2.27}), (\ref{2.28}), while
the spin-spin interaction is a non-trivial effect.

The simplest example is a straight string considered in Sec. \ref{weak1}. In this case $(a_{\mathrm{in}} \cdot b_{\mathrm{in}})=0$, 
and variation (\ref{j2.27}) can be written as 
\begin{equation}\label{j2.28}
\delta \vec{j}=-4\pi \mu G [\vec{J}\times\vec{p}]\,\mathrm{sign}(a_{\mathrm{in}}^y)~~,
\end{equation}
where $p^i=\delta^i_z$ is the unit vector along the string (directed along the $z$-axis).  

If the conservation of the total angular momentum 
is taken into account one comes to the following variation of the angular momentum of the black hole:
\begin{equation}\label{j2.29}
\delta\vec{J}=4\pi \mu G [\vec{J}\times\vec{p}]\,\mathrm{sign}(a_{\mathrm{in}}^y)~~.
\end{equation}
The momentum rotates around the string axis, the rotation angle being  determined  by the string energy only.
The sign of rotation in (\ref{j2.29}) depends on the position of the black hole with respect to the world-sheet of the string. If 
the string moves between two rotating black holes it causes a non-trivial relative rotation of their angular momenta by the angle $8\pi \mu G$.

\subsection{Scattering and holonomy of the string spacetime}\label{holon}

Results of Secs. \ref{weak1}, \ref{angm}  can be interpreted as change of the velocity and angular momentum of a black hole
in the gravitational field of a null string. Transformations of trajectories of massive particles and light rays in the gravitational 
field of the string have been studied in \cite{Fursaev:2017aap}, \cite{Fursaev:2018spa} by using holonomy of spacetime created by a null string.
It is instructive to see that Secs. \ref{weak1}, \ref{angm} reproduce results of \cite{Fursaev:2017aap} for the case of a straight null string. 
Transformation of a vector 
$V^\mu$  under a parallel transport around the string is 
\begin{equation}\label{h2.8a}
(V')^0=\left(1+{\omega^2 \over 2}\right)V^0-{\omega^2 \over 2}V^x+\omega V^y~~,
\end{equation}
\begin{equation}\label{h2.8b}
(V')^x=\left(1-{\omega^2 \over 2}\right)V^x+{\omega^2 \over 2}V^0+\omega V^y~~,
\end{equation}
\begin{equation}\label{h2.8c}
(V')^y=V^y+\omega(V^0-V^x)~~,~~(V')^z=V^z~~,
\end{equation}
where $\omega=8\pi \mu G$. As before, the string is assumed to be directed along the $z$ axis and move in the $x$ direction.
Eqs. (\ref{h2.8a})--(\ref{h2.8c}) are null transformations which are reduced to (\ref{1.6}) in
the case of the tetrade $l,n,p,q$.  

If the string moves between two massive bodies which are initially at rest,  (\ref{h2.8a})--(\ref{h2.8c}) imply 
that the bodies acquire a relative coordinate velocity
toward each other, in the direction orthogonal to the string,
\begin{equation}\label{j2.30}
v^y\simeq \omega=8\pi \mu G~~
\end{equation}
(in the limit when $\omega \ll 1$).
This result is in agreement with (\ref{2.30a}) which yields coordinate velocity of each body $|v^y_{\mathrm{body}}|\simeq 4\pi \mu G$. 

According to (\ref{h2.8a})--(\ref{h2.8c}),  in the linear in $\omega$ approximation, a parallel transport of a spin 4-vector with components  $J^0=0, J^i$, 
generates the rotation
\begin{equation}\label{j2.31}
\delta \vec{J}=\omega [\vec{J}\times \vec{p}]~~.
\end{equation}
The parameter $\omega$ is the angle  of relative rotation of two spins when the string moves between them.
Equation (\ref{j2.31}) coincides with (\ref{j2.29}).

Therefore ``holonomy variations'' of velocities and spins of test bodies 
caused by gravitational field of a null cosmic string are in complete agreement with their ``scattering variations''. Note that the scattering
data are Noether charges (\ref{2.19}) defined by the stress-energy tensor of null strings. The above analysis yields a check of 
introduced SET  (\ref{1.10}). The check is non-trivial since  ``scattering variations'' 
(\ref{2.30a}), (\ref{j2.29})  are determined by independent contributions from all segments of the string.

\subsection{Scattering and null string optics}\label{opt}   

One can also use solution (\ref{2.16c}), (\ref{2.16d}) to see how expansion and rotation 
parameters of the string at $\mathcal{I}^{-}$ and $\mathcal{I}^{+}$ are related.  For  simplicity
we assume that the strings are ``unfrozen''
at $\mathcal{I}^{-}$.  For such strings the asymptotics of 
optical scalars
at $\mathcal{I}^{-}$ and $\mathcal{I}^{+}$ are, respectively,
\begin{equation}\label{2.31a}
Z_{\mathrm{in}} (\lambda,\tau)=\frac{1}{\lambda}+\frac{z_{\mathrm{in}}}{\lambda^2}+O(\lambda^{-3})~~,
\end{equation}
\begin{equation}\label{2.31b}
Z_{\mathrm{out}} (\lambda,\tau)=\frac{1}{\lambda}+\frac{z_{\mathrm{out}}}{\lambda^2}+O(\lambda^{-3})~~.
\end{equation}
Expansion is $\theta=1/\lambda+\Re~ z/\lambda^2+...$, the rotation is  $\kappa=\Im~ z/\lambda+...$. After some algebra one gets
$$
\Re~ z_{\mathrm{out}}=\Re~ z_{\mathrm{in}}+\frac{2r_g}{r_0^4}\left[r_0^4\ln\frac{\varrho}{r_0}+
r_0^2\left((\dot{a}_{\mathrm{in}} \cdot \dot{a}_{\mathrm{in}})-(a_{\mathrm{in}} \cdot b_{\mathrm{in}})^2\right)\right.
$$
\begin{equation}\label{2.32a}
\left. +(\dot{b}_{\mathrm{in}}\cdot a_{\mathrm{in}})^2\left((a_{\mathrm{in}} \cdot b_{\mathrm{in}})^2+
a_{\mathrm{in}}^2\right)-
2(\dot{a}_{\mathrm{in}}\cdot a_{\mathrm{in}})^2\right],
\end{equation}
$$
(\Im~ z_{\mathrm{out}})^2=(\Im~ z_{\mathrm{in}})^2-\frac{4r_g}{r_0^4}\left[r_0^2
\left((\dot{a}_{\mathrm{in}}\cdot a_{\mathrm{in}})(a_{\mathrm{in}} \cdot b_{\mathrm{in}})+
(\dot{a}_{\mathrm{in}}\cdot \dot{a}_{\mathrm{in}})\Re~ z_{\mathrm{in}}\right)+\right.
$$
\begin{equation}\label{2.32b}
\left.(\dot{b}_{\mathrm{in}} \cdot a_{\mathrm{in}})\left((\dot{a}_{\mathrm{in}}\cdot a_{\mathrm{in}})+(\dot{b}_{\mathrm{in}} \cdot a_{\mathrm{in}})^2\Re~ z_{\mathrm{in}}\right)
\left(a_{\mathrm{in}}^2+(a_{\mathrm{in}} \cdot b_{\mathrm{in}})^2\right)-
2(\dot{a}_{\mathrm{in}}\cdot a_{\mathrm{in}})^2\left((a_{\mathrm{in}} \cdot b_{\mathrm{in}})+\Re~ z_{\mathrm{in}}\right)\right]~~,
\end{equation}
where $r_0=r_0(\tau)$.

Equations (\ref{2.32a}), (\ref{2.32b}) demonstrate that the string evolution in the space of optical parameters $(\theta,\kappa)$ is predictable in 
the following  sense: given the initial data
one can determine a final state of the string.

\section{Null strings in asymptotically flat space-times}\label{AFST}
\setcounter{equation}0

\subsection{Formulation of the problem and the Bondi-Sachs formalism}\label{Z}

Consider a behavior of the string scalar in asymptotically flat spacetimes near $\mathcal{I}^{+}$. 
Our aim is to see, by using string optical equation
(\ref{i.1}), how  the spacetime content is encoded in $Z$. From now on we denote $Z_{\mathrm{out}}=Z$, subscripts
``in'' and ``out'' will be omitted since we focus on effects related to features of gravitational fields rather than
on transformations of the trajectories.

It is 
convenient to use the Bondi-Sachs coordinates $x^\mu=(u,r,x^A)$, $A=1,2$, based on a family of outgoing null hypersurafces,  see e.g. \cite{Madler:2016xju}. 
The corresponding metric is a generalization of (\ref{2.21}):
\begin{equation}\label{3.1}
ds^2=-\frac{V}{r}e^{2\beta}du^2-2e^{2\beta}dudr+h_{AB}(dx^A- U^Adu)(dx^B-U^Bdu)~~.
\end{equation}
The null hypersurfaces in question are  $u=c$, where $c$ is a constant. Coordinate $r$ which varies along null rays is chosen to be areal coordinate.  
The future null  infinity is at $r\to+\infty$.

Metric (\ref{3.1})
is flat when $V=r$, $\beta=U^A=0$, $h_{AB}=r^2\gamma_{AB}$, with $\gamma_{AB}$ being a metric on a unit 2-sphere.
In an asymptotically flat space-time at large $r$ (we set Newton constant $G=1$)
\begin{equation}\label{3.2a}
\frac{V}{r} = 1 - \frac{2M}{r} + O(r^{-2})~~,
\end{equation}
\begin{equation}\label{3.2b}
h_{AB} = \gamma_{AB}+ \frac{C_{AB}}{r} +\frac{D_{AB}}{r^2} + O(r^{-3})~~,
\end{equation}
\begin{equation}\label{3.2c}
\beta =  -\frac{1}{32} \frac{C^{AB}C_{AB}}{r^2} +O(r^{-3})~~,
\end{equation}
\begin{equation}\label{3.2d}
U^A =-\frac{\eth_B C^{AB}}{2 r^2} + \frac{1}{r^3}\left( 2J^A + \frac{1}{3} C^{AE} \eth^F C_{EF} \right)+
O(r^{-4})~~,
\end{equation}
\begin{equation}\label{3.2e}
e^{2\beta} = 1- \frac{1}{16} \frac{C^{AB}C_{AB}}{r^2} +O(r^{-4})~~.
\end{equation}
Here $\gamma_{AB}=\mathrm{diag}(1,\sin^2\theta)$ is the metric on unit sphere $S^2$, $\eth_A$ 
is a covariant derivative on $S^2$.  Indices $A,B$ in (\ref{3.2c})-(\ref{3.2e}) are raised and lowered 
with the help of $\gamma_{AB}$. One can show that \cite{Madler:2016xju}
\begin{equation}\label{3.3}
D_{AB}=\frac 14 \gamma_{AB} C^{DE}C_{DE}~~.
\end{equation}
We denote the set of coordinates $x^A$ on $S^2$ by $\Omega$.
Quantities $M=M(u,\Omega)$ and $J^A=J^A(u,\Omega)$ are the mass and the angular momentum aspects, respectively.  
$C_{AB}=C_{AB}(u,\Omega)$ is a traceless tensor (the strain) on a tangent space to $S^2$. The term 
${C_{AB} / r}$ in (\ref{3.2b})  is a perturbation of the metric caused by the outgoing gravitational radiation.
One also defines Bondi news tensor $N_{AB} = \partial_u C_{AB}$. If the vacuum Einstein equations are satisfied, 
\begin{equation}\label{3.2g}
2 \partial_u M =\eth_A \eth_B N^{AB} - N_{AB} N^{AB}~~.
\end{equation}
A similar relation can found for $\partial_u J_A$.

The optical scalar of a string in a flat spacetime has the following asymptotic, see
(\ref{2.4}):
\begin{equation}\label{3.4}
Z(\lambda,\tau)\sim \sum_{k=1}(-1)^k z^k(\tau)~ \lambda^{-k}~~.
\end{equation}
In arbitrary spacetime near $\mathcal{I}^{+}$, according to the peeling theorem \cite{Adamo:2009vu},
\begin{equation}\label{3.5a}
\Psi_0 \bigl|_{\lambda \to \infty} = \psi^0_0(\tau)\lambda^{-5} + O(\lambda^{-6})~~~,
\end{equation}
\begin{equation}\label{3.5b}
\Phi_{00} \bigl|_{\lambda \to \infty} = \phi^0_{00}(\tau)\lambda^{-6} + O(\lambda^{-7})~~~.
\end{equation}
Therefore we look for asymptotic solution to (\ref{i.1}) as a series
\begin{equation}\label{3.6}
Z(\lambda,\tau)\sim \sum_{k=1}z_k(\tau)~ \lambda^{-k}~~.
\end{equation}	
The structure of the first coefficients in (\ref{3.6}) follows from (\ref{i.1}), (\ref{3.5a})
\begin{equation}\label{3.7}
z_1=1~~,~~z_2(\tau)=-z(\tau)~~,~~z_3(\tau)=z^2(\tau)~~,~~z_4(\tau)=-z^3(\tau)+\frac 12 \psi^0_0(\tau)~~.
\end{equation}
Our aim in next Sections is to understand how $z_k(\tau)$ depend on spacetime characteristics
$C$, $M$ and $J$.

\subsection{Leading terms in $Z$}\label{BSF}

We begin with the string equations (\ref{2.1}) in flat spacetime in Bondi-Sachs coordinates (\ref{2.21})
$u,r,\Omega$. When $\mathcal{I}^{+}$ is approached, points of string move as almost radial 
light rays,
\begin{equation}\label{3.8}
u=t-r=0~~,~~r=\lambda~~,~~x^A=x^A(\tau)~~. 
\end{equation}
This approximation is enough to calculate $z(\tau)$ in (\ref{3.7}), and, therefore, first 
coefficients  $z_2(\tau)$, $z_3(\tau)$  in asymptotic (\ref{3.6}). On the string trajectory we define functions
\begin{equation}\label{3.9}
M(\tau)=M(0,\Omega(\tau))~~,~~J(\tau)=J(0,\Omega(\tau))~~,~~C_{AB}(\tau)=C_{AB}(0,\Omega(\tau))~~.
\end{equation}
The connecting vector is $\eta=\eta^A\partial_A$, where $\eta^A=\dot{x}^A$.

We note that for (\ref{3.8}) the non-vanishing components of $p,q,m$ are
$$
p^A=\bar{p}^A/\lambda+O(\lambda^{-2})~~,~~q^A=\bar{q}^A/\lambda+O(\lambda^{-2})~~,~~
m^A={1 \over \sqrt{2}}(p^A+iq^A)=\hat{m}^A/\lambda+O(\lambda^{-2})~~,
$$
Vectors $\bar{p}^A$, $\bar{q}^A$ are unit tangent vectors on $S^2$. Therefore, one can use decompositions
\begin{equation}\label{3.10}
\gamma_{AB}=\hat{m}_A \hat{\bar{m}}_B+\hat{\bar{m}}_A\hat{m}_B~~,
\end{equation}  
\begin{equation}\label{3.11}
C_{AB}=\hat{m}_A \hat{m}_B~ \bar{C}+ \hat{\bar{m}}_A \hat{\bar{m}}_B~  C~~,
\end{equation}
\begin{equation}\label{3.12a}
C(\tau)=C_{\oplus}(\tau) +iC_{\otimes}(\tau)~~.
\end{equation}
Unkown function $z(\tau)$ can be found by calculating the corresponding spin coefficients
\begin{equation}\label{3.12}
z(\tau)=z_0(\tau)+\frac 12C(\tau) + M(\tau)~~.
\end{equation}
Quantity $z_0(\tau)$ is determined by the string trajectory in flat spacetime. To calculate $z_0(\tau)$ one would need to go beyond approximation (\ref{3.8}), but the result is already known from (\ref{2.4}).
Coefficients $C_{\oplus}$ and $C_{\otimes}$ correspond to ``$\oplus$'' and ``$\otimes$'' polarizations of gravity waves in the given basis. Their contribution to (\ref{3.12}) has been found in \cite{Fursaev:2021xlm}.

\subsection{Subleading term in $Z$}\label{subl}

According to (\ref{3.7}) first 4 coefficients $z_k$ are known if $\psi^0_0(\tau)$ in expansion
(\ref{3.5a}) of the Weyl tensor is known. To calculate $\psi^0_0(\tau)$ we need Bondi-Sachs asymptotic
(\ref{3.2a})-(\ref{3.2e}) for the metric (which yields components $C_{\mu\nu\lambda\rho}$). The basis 
vectors, $l$ and $m$, in the given approximation can be taken as in flat spacetime. However, 
to calculate the components of $l$, $m$ one should go beyond radial string approximation (\ref{3.8}).
Let us introduce a unit vector $\vec{N}$, orthogonal to $S^2$, which sets spherical coordinates on $S^2$
\begin{equation}\label{3.13}
N_1(\Omega)=\sin \theta \cos\varphi~~,~~N_2(\Omega)=\sin \theta \sin\varphi~~,~~N_3(\Omega)=\cos \theta~~.
\end{equation}
We also use three unit orthonormalized vectors
\begin{equation}\label{3.14}
\vec{e}_1=\frac{\dot{\vec{b}}}{|\dot{b}|}~~,~~ \vec{e}_3=\vec{b}~~,~~ \vec{e}_2=\left[\vec{e}_3 \times \vec{e}_1\right]~~ 
\end{equation}
and define complex scalars
\begin{equation}\label{3.15}
N= \frac{N_1 + i N_2}{\sqrt{2}}~~,~~ A = \frac{a_1 + i a_2}{\sqrt{2}}~~,~~ 
\end{equation}
where $N_i=(N \cdot e_i)$, $a_i=(a \cdot e_i)$.
With the help of (\ref{3.15}) the leading terms in asymptotics of the $l$, $m$ for a generic trajectory of a null string in 
flat spacetime in Bondi-Sachs coordinates  can be written as
\begin{equation}\label{3.16}
l^\nu = \frac{\hat{l}^\nu}{\lambda^2} +  O(\lambda^{-3})~~,~~\nu=u,\theta,\varphi~~,~~
l^r = 1-\frac{\hat{l}^r}{\lambda^2} +  O(\lambda^{-3})~~,	
\end{equation}
\begin{equation}\label{3.18}
\hat{l}^u=\hat{l}^r=|A|^2~~,~~\hat{l}^\theta=\frac{1}{\sin\theta} 2 \Re\left( N \bar{A} \right)~~,~~
\hat{l}^\varphi=\frac{1}{\sin^2\theta} 2 \Im\left( \bar{N} A \right)~~,
\end{equation}
\begin{equation}\label{3.17}
m^\nu = \frac{\hat{m}^\nu}{\lambda}  +  O(\lambda^{-2})~~,
\end{equation}
\begin{equation}\label{3.19}
\hat{m}^u = -A~~,~~\hat{m}^r =A~~,~~\hat{m}^\theta = -\frac{N}{\sin\theta} ~~,~~~
\hat{m}^\varphi = -i\frac{N}{\sin^2\theta}~~.
\end{equation}
The computation of the leading asymptotic of the Weyl tensor (\ref{3.5a}) is rather lengthy. It yields
$$
\psi^0_0=12 A~ \hat{m}_B J^B	+6A^2~ M 	
+\frac{1}{2} A^4 ~\hat{\bar{m}}_A \hat{\bar{m}}_B  \partial_u   N^{AB}
-2 A^3   \hat{\bar{m}}_B \eth_A N^{AB}
$$
$$
+\frac{3}{2} A^2 \left(\hat{m}_A \hat{\bar{m}}_B \gamma_{EF} C^{AE}   N^{BF} 
+\eth_A \eth_B C^{AB}
- 2 \hat{\bar{m}}^B \hat{m}_E \eth_B \eth_A C^{AE}\right)
$$
\begin{equation}\label{3.20}
-A \left( \hat{m}^A C_{AB} \eth_E C^{EB} + \frac{3}{8} \hat{m}_A \eth^A \left( C^{BD} C_{BD} \right) \right)
-\frac{3}{8} \hat{m}_A \hat{m}_B C^{AE} C^{BF} C_{EF}~~.
\end{equation}
Indices in (\ref{3.20}) are raised and lowered with the help of $\gamma_{AB}$. 

The first two terms in the right hand side (r.h.s.) of (\ref{3.20}) represent the mass and 
angular aspects of the spacetime. They yield contribution
both to parameters of asymptotic expansion $\theta=\Re ~Z$ and asymptotic rotation $\kappa=\Im~ Z$ of the string,
see (\ref{3.6}), (\ref{3.7}). The angular momentum aspect $J^A$
appears only at the order $\lambda^{-4}$. According to (\ref{3.7}), (\ref{3.12}) the mass aspect $M$ appears at the order $\lambda^{-2}$, where it contributes  only to $\theta$.

The terms in  (\ref{3.7}), (\ref{3.20}) which depend on the strain $C_{AB}$ and its covariant derivatives describe contributions to $Z$ from the gravitational background.  The Bondi news tensor $N_{AB}$ which determines the energy flux across  $\mathcal{I}^{+}$, see 
(\ref{3.2g}), appears at the order $\lambda^{-4}$.

Some comments on special cases are in order. Sec. \ref{BSF} describes 
null strings which move to $\mathcal{I}^{+}$ along radial light rays. For such strings, vector $m^\mu$  has vanishing components $m^u=m^r=0$, see 
equations of motion  (\ref{3.8}) . Equations
(\ref{3.19}) imply that $A=0$, and $\vec{a}$ is directed along $\vec{b}$. Then $\psi^0_0$ is reduced to the last term in the r.h.s. 
Therefore, radial strings are not sensitive, at the order $\lambda^{-4}$,  to the mass and angular aspects, as well as to the energy flux.

Another case are null strings which are ``asymptotically frozen'', that is $\dot{b}=0$, see Sec. \ref{flat}, and
$a_2=\Im~A=0$. According to  (\ref{3.20}) mass and angular aspects do not contribute to the rotation parameter $\kappa$
of such strings at the order $\lambda^{-4}$.

\subsection{Asymptotics of string energy}\label{energyas}

Energy evolution equation (\ref{i.2}) and Eqs. (\ref{3.6}),  (\ref{3.20}) allow one to 
find important asymptotics of the string energy at $\mathcal{I}^{+}$: 
\begin{equation}\label{3.21}
\mu(\lambda,\tau)\sim \mu(\tau)~\sum_{k=1}f_k(\tau)~ \lambda^{-k}~~,
\end{equation}
\begin{equation}\label{3.22}
f_1=1~~,~~f_2=-\Re~z~~,~~f_3=\frac 12\left[\Re~z^2+(\Re~z)^2\right]~~,
\end{equation}
\begin{equation}\label{3.23}
f_4=-\frac 16 \left[(\Re~z)^3+3\Re~z~ (\Re~z)^2+2\Re~z^3-\Re~\psi^0_0\right] ~~.
\end{equation}
Here $\mu(\tau)$ is a coefficient which cannot be determined by   (\ref{i.2}). Parameters $z(\tau)$, $\psi^0_0$ are given by (\ref{3.12}),
 (\ref{3.20}), respectively.

Equations  (\ref{3.21})-(\ref{3.23}) demonstrate optical properties of null strings when space-time content
determines the evolution of their physical energy $\mu(\lambda,\tau)$. 
In the cosmological context, null strings are similar to relic photons whose energy encodes properties of the Universe they travel.

\section{Discussion}\label{sum}

The origin of null strings may be related to physics of
fundamental strings at Planckian energies \cite{GM1}, \cite{GM2}, \cite{Lee:2019wij}, \cite{Xu:2020nlh}.
If fundamental null strings were produced in the early Universe they might be stretched to cosmological scales and become
cosmic strings, see \cite{Sarangi:2002yt}, \cite{Copeland:2003bj} for discussion of such scenario  for fundamental tensile strings.  

If null cosmic strings exist,  it is important to describe observable physical effects they may produce. It is also important 
to understand specific features of null cosmic strings which distinguish them from the tensile cosmic strings. 
Since velocities of astrophysical tensile cosmic strings are expected to be below 0.7 of the speed of light, interactions of null and tensile cosmic
strings with black holes look different, see discussion in Sec. \ref{weak1}. As a next step, it would be interesting to compare interactions 
of null and tensile strings with black holes in the regime of the strong gravity where tensile cosmic strings exhibit specific chaotic behavior, see e.g. \cite{Larsen:1993nt}, \cite{Frolov:1999pj}.

World-sheets of tensile cosmic strings may have some luminal points which can 
be located, for  example, at cusps developed by oscillating loops \cite{Vilenkin:2000jqa}.
The cusps are known to emit strong beams of high-frequency gravitational waves \cite{Damour:2000wa}.  
These gravitational waves produced at different epochs form a stochastic gravitational background. Experimental evidence of the background is 
being actively
searched for by the Advanced LIGO and Virgo Collaborations \cite{LIGOScientific:2021nrg}.

The key feature of null strings is that they behave as one-dimensional null geodesic congruences and are characterized 
by optical parameters $\theta$ and $\kappa$. In the present paper we extended the optical analogy of null strings. We introduced
the stress-energy tensor of a null string and, with its help, gave the definition of the physical energy $\mu$ of the string per unit length.
As was shown, null strings develop caustics which accumulate large amounts of energy. We expect that caustics of null strings, similar to cusps, may emit 
gravitational waves which contribute to the gravitational background. Studying these effects is in progress.

Null cosmic strings  may carry an important information about the spacetime content and the physical processes in the early Universe. 
The explicit dependence of $\mu$ on the strain and Bondi news tensors of gravitational wave background, mass and angular momentum aspects 
has been obtained near the  future null infinity in an analytic form in asymptotically flat spacetimes. An intriguing property of null cosmic strings is that
they, like relic photons,
may encode the story of their propagation in the Universe.

\section{Acknowledgments}

This research is supported by Russian Science Foundation grant No. 22-22-00684, \\
https://rscf.ru/project/22-22-00684/.

\end{document}